# OUT-OF-DISTRIBUTION MULTI-VIEW AUTO-ENCODERS FOR PROSTATE CANCER LESION DETECTION

*Alvaro Fernandez-Quilez*[1,2,‡]
*Linas Vidziunas*[1,2,†]   *Ørjan Kløvfjell Thoresen*[1,2,†]
*Ketil Oppedal*[1]   *Svein Reidar Kjosavik*[3]   *Trygve Eftestøl*[1]

[1]Department of Electrical Engineering and Computer Science, University of Stavanger, Norway.
[2] SMIL, Department of Radiology, Stavanger University Hospital, Norway.
[3] General Practice and Care Coordination Research Group, Stavanger University Hospital, Norway.

## ABSTRACT

Traditional deep learning (DL) approaches based on supervised learning paradigms require large amounts of annotated data that are rarely available in the medical domain. Unsupervised Out-of-distribution (OOD) detection is an alternative that requires less annotated data. Further, OOD applications exploit the class skewness commonly present in medical data. Magnetic resonance imaging (MRI) has proven to be useful for prostate cancer (PCa) diagnosis and management, but current DL approaches rely on T2w axial MRI, which suffers from low out-of-plane resolution. We propose a multi-stream approach to accommodate different T2w directions to improve the performance of PCa lesion detection in an OOD approach. We evaluate our approach on a publicly available data-set, obtaining better detection results in terms of AUC when compared to a single direction approach (73.1 vs 82.3). Our results show the potential of OOD approaches for PCa lesion detection based on MRI.

## 1. INTRODUCTION

The deep learning (DL) field convergence with medical imaging has enabled the development of tools that hold the potential to automatize tasks that otherwise would be prone to be time-intensive, subject to inter-reader variability and overall, sub-optimal in the absence of specialized training [1, 2]. To date, the main learning paradigm for DL-powered applications has been supervised learning [3], which relies on large amounts of annotated data. The acquisition of such annotations can be particularly challenging in the medical domain where specialized knowledge and privacy concerns are a barrier to the access and acquisition of annotated data [3, 4].

† Equal contribution
‡ Corresponding author: alvaro.fernandez.quilez@sus.no

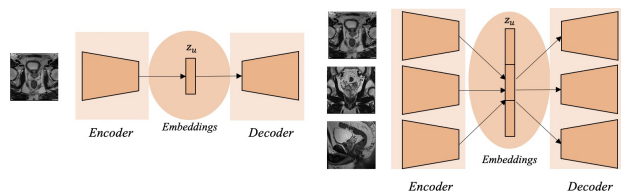

**Fig. 1**. Single view (left side) and multi-view AE (right side) for OOD-based PCa lesion detection.

Specifically, MRI can help with the management and diagnostic pathway of diseases such as PCa [5, 6] by, for instance, increasing the detection rate of lesions that require further testing [7].

Annotation scarcity is usually tackled by exploiting ImageNet weights and transfer learning. In spite of it, the gap between the initialization domain and the medical one can lead to sub-optimal results [8]. In addition, current medical-imaging self-supervised learning (SSL) [9] approaches are not tailored to tackle the data challenges present in medical imaging classification such as *class skewness*.

Class skewness is a prevalent issue where the number of normal controls present in the data-set outnumber the cases that present a pathology, potentially biasing the training of the architecture towards maximizing (or minimizing) the metric of choice in terms of the over-represented class [10]. In that regard, OOD approaches present an alternative to SSL and supervised DL approaches [11]. Specifically, the OOD approach exploits the prevalence of normal controls by learning their distribution after seeing a handful of them, mimicking, roughly speaking, the way a health professional would learn to discern between normal controls and cases with a pathology [12].

Existing OOD attempts to PCa lesion detection have relied on axial T2w MRI, which suffers from low out-of-plane resolution. In addition, sagittal and coronal MRI scans are commonly available at the time of acquisition: they are typically acquired by default and taken into consideration by radiologists when carrying out the analysis [13]. In this work, we propose a multi-view variational auto-encoder (mVAE) able to accommodate the orthogonal MRI directions acquired by default for PCa. We formulate the problem from an OOD perspective where our proposed mVAE exploits the large number of controls (*no lesion*) present in the data-set. Our contributions can be summarized as follows:
- We investigate OOD as a way to mitigate the lack of annotated data and class-imbalance in PCa lesion detection based on MRI.
- We propose a *multi-view* OOD approach and compare it with a *single-view* (axial-based) one (Figure 1) in terms of different metrics and 95 % confidence intervals (CI) (Figure 2).
- We explore different ways of obtaining the final lesion detection: *threshold* and *fine-tuning*-based (Figure 3).

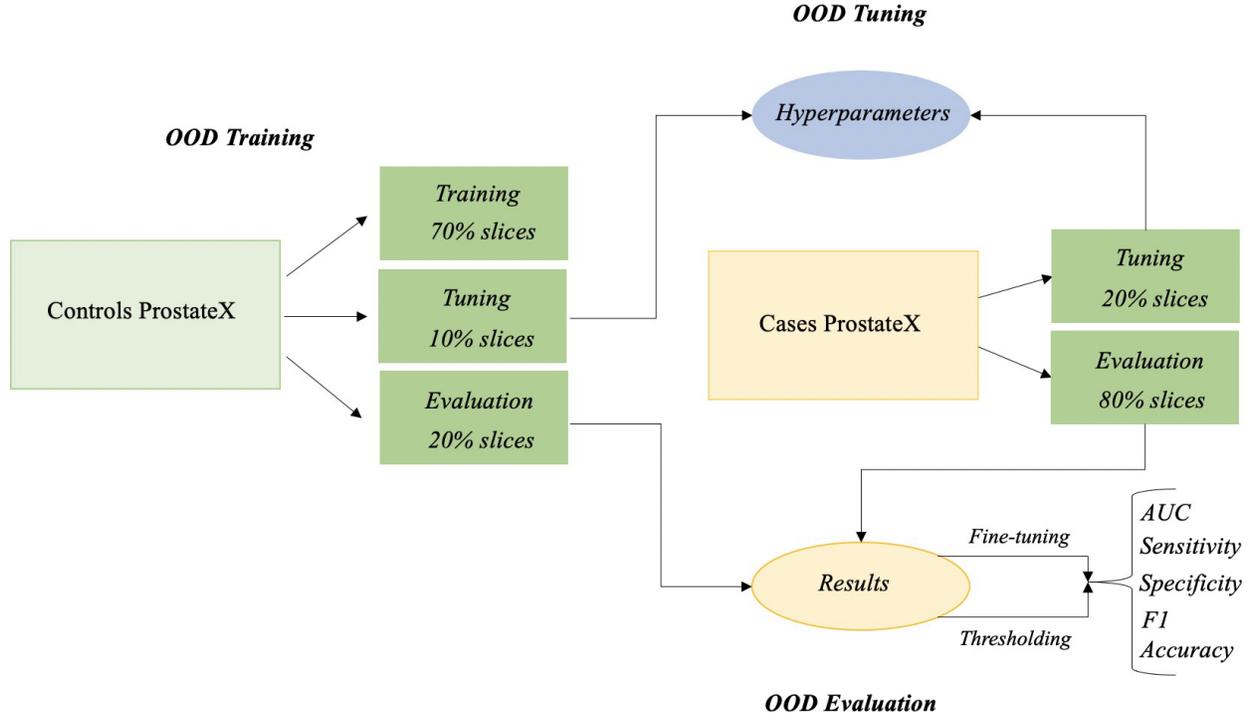

**Fig. 2**. Out-of-distribution single view and multi-view technical approach: training and evaluation.

## 2. METHODS

### 2.1. Data

We use the open-access data-set ProstateX [14]. The study has a retrospective nature and different MRI sequences are included, from which we focus on T2w axial, sagittal and coronal MRI. The sequences are acquired with a 3.0 T field strength Siemens scanner without an endorectal coil. Specifically, the cohort included in the study consists of *204 patients* which translates into *330 lesions* and a total of *12845 2D slices*, where *1109 contain a lesion* (≈9% of the total amount of slices, which clearly depicts class skewness).

The data used to develop the study uses the publicly available data along with the ground truth (training set), at the time this study is carried out. The 204 sequences are provided with a biopsy result which are used to obtain the labels (*ground truth*) for each sequence. We label in a binary fashion each slice (2D) depending on whether a lesion is present or not in the slice under consideration.

#### 2.1.1. Pre-processing and data splitting

We start by re-sampling the different T2w sequence directions by linear interpolation to a common coordinate system with a resolution of 0.5x0.5x3.6mm, which resembles the predominant 2D resolution and slice thickness among the different sequence directions. Following, the pixel intensity of the sequences is normalized to a range of [0, 1] and outlier removal is applied by forcing the pixel intensity of each sequence to lie between the 1st and 99th pixel intensity percentile.

In order to train, tune and evaluate the model we start by dividing the available slices in controls (*no lesion*) and cases (*lesion is present*). Following, we divide the controls in 70%/20%/10% for train, tuning and evaluation, respec-tively. A similar process is followed for the cases but omitting the training split, since we only make use of normal control cases at training time. Hence, the cases are divided in 80%/20% for evaluation and tuning purposes. The data splitting is performed in a stratified way in both cases, avoiding cross-contamination in the form of data leakage (i.e. slices of the same patient present in the different sets). Figure 2 depicts the technical approach to the project.

**2.2. OOD detection with auto-encoders**

Our OOD detection framework is based on *Variational Auto-Encoders (VAE)*[1]. Generally speaking, OOD detection based on VAE models consists of approximating the likelihood of a data point with respect to the distribution they were trained on. We approach the OOD detection from two points of view: *single-view* and *multi-view* (Figure 1).

*2.2.1. Single-view*

In our first approach, single-view VAE (sVAE), we train the VAE exclusively with a set of $N$ T2w axial MRI slices (controls) $X_h = (x_{h_1}, x_{h_2}..., x_{h_N})$ (Figure 1, illustration on the left side). Our ultimate goal is to learn the distribution $P(X_h)$ by training the VAE to *reconstruct* axial control slices. In the evaluation stage, we exploit the prior distribution to determine the probability of an axial T2w slice (either case or control) with respect to the distribution of the learnt $P(X_h)$. The underlying rationale is that we expect the model to not be able to reconstruct the cases presenting a pathology $X_a = (x_{a_1}, x_{a_2}..., x_{a_N})$ accurately, which will indicate a low probability with respect to $P(X_h)$.

*2.2.2. Multi-view*

In the multi-view approach (mVAE), we follow similar steps to the sVAE approach. In mVAE, we integrate the different orthogonal prostate MRI views (axial, coronal and sagittal) by means of an early concatenation (Figure 1, right side). Specifically, we use three independent encoders that are trained, respectively, with a set of $N$ T2w axial MRI slices $X_{h_a}$, T2w coronal slices $X_{h_c}$ and T2w sagittal slices $X_{h_s}$. Each encoder is trained to learn the distribution of the respective views' controls. Following, we combine the representations learnt by each of the encoders before the bottleneck layer of the VAE architecture.

In the decoding phase, the merged representations are split again into independent representations to produce individual reconstructions of the different prostate MRI view inputs (Figure 3). During training, we aim to minimize an objective function $L_T$ that is the result of a weighted combination of the loss functions for each MRI view:

$$L_T = \lambda L_a + \delta L_c + \gamma L_s \quad (1)$$

where $\lambda L_a$, $\delta L_c$ and $\gamma L_s$ are the loss functions and tunable weight parameters for the axial view, coronal view and sagittal view, respectively. Following, we exploit the prior control distributions learnt by the encoders to determine the probability of a T2w slice (case or control) with respect to $P(X_h)$. Nevertheless, at evaluation time we use, exclusively, *axial data* under the assumption that is the plane of reference and the other two serve to improve training and the final detection results.

[1]https://github.com/LinasVidziunas/Unsupervised-lesion-detection-with-multi-view-MRI-and-autoencoders

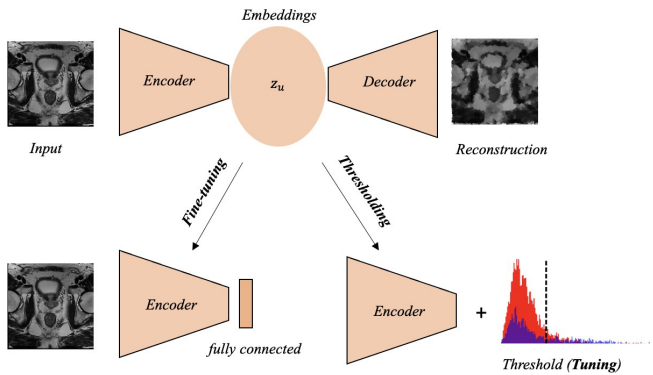

**Fig. 3**. Out-of-distribution single view and multi-view detection: fine-tuning (left) and threshold-based (right).

*2.2.3. Fine-tuning and thresholding-based OOD*

During evaluation, we explore two different approaches for both sVAE and mVAE: *fine-tuning* and *thresholding*-based lesion detection (Figure 2). Specifically, for the fine-tuning approach we draw inspiration from other works where a pre-text task is defined and later, the architecture are fine-tuned on a different final task (downstream) [2, 15]. In order to perform the *fine-tuning* evaluation we proceed in the following way: We discard the decoder part of the VAE architecture and use the encoder with the weights obtained from training the VAE with the reconstruction objective and using normal controls. Following, we add a fully connected layer and fine-tune the whole encoder while changing the loss function to a cross-entropy one (classification).

In the case of the *thresholding*-based evaluation, we make use of the reconstruction loss obtained for each input and find a threshold by means of the tuning set and the calculation of the inter-quartile range (IQR) [11, 16], which does not assume any underlying distribution for the reconstruction loss distributions. Finally, the obtained threshold is used during testing to discriminate and obtain a hard-thresholded classification of each slice. Figure 3 illustrates in from a schematic perspective the way both approaches work.

**Table 1**. Results for sVAE and mVAE with the best OOD detection method: fine-tuning

| Method | Accuracy | Precision | Recall | F1 | macro-AUC | p value |
|---|---|---|---|---|---|---|
| sVAE | 78.2 [78.0, 78.4] | 74.1 [73.8, 74.4] | 65.6 [64.7, 66.5] | 69.6 [68.9, 70.2] | 73.1 [72.9, 73.3] | <0.001† |
| mVAE | 83.9 [83.6, 84.2] | 77.8 [77.6, 78.0] | 70.0 [69.5, 70.5] | 73.7 [73.3, 74.0] | **82.3** [81.8, 82.8] | - |

† Statistically significant.

Both architectures are trained for 250 epochs with an early stopping criteria if the validation loss did not improve for 30 consecutive epochs, an Adam optimizer, a learning rate of $1e^{-4}$, a batch size of 32 and an NVIDIA V100 GPU.

### 2.3. Evaluation

Our evaluation is based on the *lesion detection ability* of the VAE architectures (sVAE and mVAE). The results are presented in terms of mean and standard deviation for macro-AUC, sensitivity, specificity, F1 and accuracy. Macro-AUC is the metric that is assumed to be the reference in terms of determining the quality of the results of the study and hence, statistical testing is based on it. We find the motivation to choose macro-AUC on the necessity to avoid skewness and weight both classes equally when evaluating the OOD frameworks.

#### 2.3.1. Bootstrapping and statistical significance

As part of the evaluation protocol, we include a non-parametric bootstrap approach with $n = 100$ replicates from the test set to estimate the variability of the model performance. Following, we obtain the 95% bootstrap confidence intervals (CI) and assess the significance at the $p = 0.05$ based on the macro-AUC. The comparison is carried out by means of Wilcoxon signed-rank [17].

### 3. RESULTS

We start by exploring the effect of threshold-based and fine-tuning based approaches for both sVAE and mVAE. As portrayed in Figure 4, we obtain a higher macro-AUC for both sVAE (73.1 [71.1, 75.1] vs 62.3 [66.3, 58.3], $p < 0.001$) and mVAE (82.3 [77.3, 87.3] vs 72.7 [66.2, 79.2], $p < 0.001$) when compared to the thresholding-based approach. Hence, we compare sVAE and mVAE on the basis of the fine-tuning evaluation protocol (Section 2.3). As shown in Table 1, the mVAE evaluated solely on axial T2w prostate MRI shows statistically significant improvements in terms of macro-AUC (73.1 [72.9, 73.3] vs 82.3 [81.8, 82.8] $p < 0.001$). In addition, we can observe improvements in the rest of the metrics as well, depicting, overall, a reduction in the number of false positives and false negatives as shown by the precision, recall and F1 metrics for the same classification threshold.

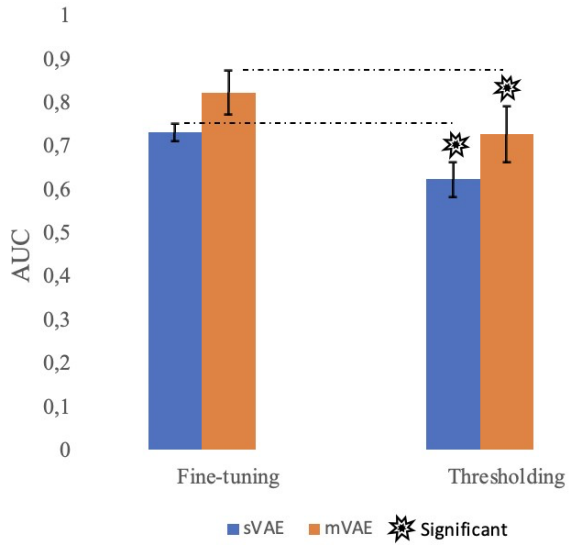

**Fig. 4**. Bootrsapped AUC and CI intervals for sVAE and mVAE in fine-tuning and thresholding evaluation approaches.

### 4. CONCLUSIONS

Our results show that the inclusion of the different orthogonal directions improves PCa lesion detection results obtained on the direction of reference: axial. In addition, our work shows the potential of OOD approaches for PCa lesion detection based on MRI. To the best of our knowledge, this work is the first to highlight the benefit of incorporating the different prostate MRI orthogonal directions that are commonly acquired by default in an OOD framework. Limitations of the work include but are not limited to the lack of a prospective validation, external validation to ensure the generalization of our results and the retrospective nature of the data used for the study.

### 5. COMPLIANCE WITH ETHICAL STANDARDS

This research study was conducted retrospectively using human subject data made available in open access by the *Prosta-*

*teX* challenge organizers. Ethical approval was not required as confirmed by the license attached with the open access data.

## 6. ACKNOWLEDGMENTS

The work has been funded by HelseVest research fundings (Stavanger University Hospital). The authors have no relevant financial or non-financial interests to disclose.